\documentclass[12pt,a4paper]{article}
\usepackage{epsf,amsfonts,amssymb}

\newcommand{\eqn}[1]{(\ref{#1})}
\newcommand{\be}{\begin{equation}}
\newcommand{\ee}{\end{equation}}
\newcommand{\ben}{\begin{displaymath}}
\newcommand{\een}{\end{displaymath}}
\newcommand{\bea}{\begin{eqnarray}}
\newcommand{\eea}{\end{eqnarray}}
\newcommand{\bean}{\begin{eqnarray*}}
\newcommand{\eean}{\end{eqnarray*}}
\newcommand{\nn}{\nonumber \\}
\newcommand{\ba}{\begin{array}}
\newcommand{\ea}{\end{array}}
\newcommand{\bi}{\begin{itemize}}
\newcommand{\ei}{\end{itemize}}

\def\a {\alpha}

\newcommand{\calf}{\mbox{${\cal F}$}}

\newcommand{\caln}{\mbox{${\cal N}$}}

\newcommand{\mbf}[1]{\mbox{\boldmath ${#1}$}}

\newcommand{\bbe}[1]{{\mathbb E}^{#1}}

\newcommand{\fc}{\frac}

\newcommand{\sac}{\, , \qquad}

\newcommand{\sect}[1]{\setcounter{equation}{0}\section{#1}}

\newcommand{\jhep}[3]{{J. High Energy Phys.} {\bf #1} {(#2)} #3}

\newcommand{\npb}[3]{{Nucl. Phys.} {\bf B #1} {(#2)} #3}

\newcommand{\prd}[3]{{Phys. Rev.} {\bf D #1} {(#2)} #3}
\newcommand{\prep}[3]{{Phys. Rep.} {\bf #1} {(#2)} #3}
\newcommand{\prl}[3]{{Phys. Rev. Lett.} {\bf #1} {(#2)} #3}

\newcommand{\hepth}[1]{{\tt hep-th/#1}}

\begin{document}

\begin{titlepage}

\bigskip
\rightline{}
\rightline{DAMTP-2002-32}
\rightline{\hepth{0204062}}
\bigskip\bigskip\bigskip\bigskip
\centerline{\Huge \bf {Supercurves}}
\bigskip\bigskip
\bigskip\bigskip

\centerline{\large \bf {David Mateos${}^1$, Selena Ng${}^2$ and
Paul K. Townsend${}^3$}}
\bigskip

\centerline{\em Department of Applied Mathematics and Theoretical Physics,}
\centerline{\em Centre for Mathematical Sciences,}
\centerline{\em Wilberforce Road, Cambridge CB3 0WA, United Kingdom}
\medskip
\bigskip\bigskip

\begin{abstract}
The TST-dual of the general 1/4-supersymmetric D2-brane supertube is
identified as a 1/4-supersymmetric IIA `supercurve': a string with
arbitrary transverse displacement travelling at the speed of light.
A simple proof is given of the classical upper bound on the
angular momentum, which is also recovered as the semi-classical limit of
a quantum bound. The classical bound is saturated by a `superhelix',
while the quantum bound is saturated by a bosonic oscillator state
in a unique $SO(8)$ representation.
\end{abstract}

\vfill

\footnoterule
{\footnotesize ${}^1$D.Mateos@damtp.cam.ac.uk \vskip -12pt}
\vskip 10pt
{\footnotesize ${}^2$S.K.L.Ng@damtp.cam.ac.uk \vskip -12pt}
\vskip 10pt
{\footnotesize ${}^3$P.K.Townsend@damtp.cam.ac.uk \vskip -12pt}

\end{titlepage}

%%%%%%%%%%%%%%%%%%%%%%%%%%%%%%%%%%%%%%%%%%%%%%%%%%%%%%%%%%%%%%%%%%%%%%%%%%%%
%%%%%%%%%%%%%%%%%%%%%%%%%%%%%%%%%%%%%%%%%%%%%%%%%%%%%%%%%%%%%%%%%%%%%%%%%%%%
%%%%%%%%%%%%%%%%%%%%%%  SECTION 1 %%%%%%%%%%%%%%%%%%%%%%%%%%%%%%%%%%%%%%%%%%
%%%%%%%%%%%%%%%%%%%%%%%%%%%%%%%%%%%%%%%%%%%%%%%%%%%%%%%%%%%%%%%%%%%%%%%%%%%%
%%%%%%%%%%%%%%%%%%%%%%%%%%%%%%%%%%%%%%%%%%%%%%%%%%%%%%%%%%%%%%%%%%%%%%%%%%%%
%\newpage
\sect{Introduction}
\label{intro}
%%%%%%%%%%%%%%%%%%%%%%%%%%%%%%%%%%%%%%%%%%%%%%%%%%%%%%%%%%%%%%%%%%%%%%%%%%%

A supertube is a 1/4-supersymmetric tubular D2-brane configuration
of IIA superstring theory that is supported against collapse by
the momentum generated by crossed electric and magnetic Born-Infeld (BI)
fields on the D2-brane \cite{MT,EMT}. As originally described, the tube
was assumed to have a circular cross-section, in which case the stability
is easily understood as a result of the centrifugal force due to an
angular momentum 2-form on the plane of the circle. The T-dual (along
the axis of the cylinder) of this rotationally-invariant supertube is a
1/4-supersymmetric helical IIB D-string \cite{CO} supported from collapse
by the combination of angular momentum and momentum along the D-string.
The TST-dual is an analogous IIA helical string \cite{MT}.

Following \cite{BK}, in which it was shown that a supertube cross-section
could be elliptical rather than circular, we have recently shown that 1/4
supersymmetry actually allows a  supertube to have a cross-section that
is an {\it arbitrary} curve in $\bbe{8}$ \cite{MNT}. Here we show that this
rather surprising fact becomes less surprising once one considers the
TST-dual configuration, which we call a `supercurve': it turns out that a
periodic supercurve is equivalent to a IIA string carrying a transverse
wave of arbitrary profile at the speed of light, and it has been known for
some time that such configurations preserve 1/4
supersymmetry \cite{DGHW,LM}.

Although a periodic `supercurve' is equivalent to a wave on a string,
this is not the description of it that emerges directly from a TST-duality
of the supertube. The two features of the supertube that are essential to
its stability, and 1/4 supersymmetry,  are (i) that the BI
electric field take the value that would be `critical' in the absence of a
magnetic field, and (ii) that the BI magnetic field not change sign. For the
TST-dual IIA string, (i) becomes a condition of uniform motion of the
string at the speed of light along the T-dual direction, while (ii) becomes
the condition that the shape of the curve be monotonic in this direction.
One might think that a string that moves uniformly in one direction at the
speed of light must have a vanishing mass density. However, because a
Nambu-Goto string is boost-invariant, only the orthogonal velocity
component of a string element is physical and this ensures that its
physical velocity is subluminal. Here one can see that an inconsistency
could arise if the monotonicity condition were violated because this would
allow the string to close, but if a closed string in $\bbe{9}$ moves at the
speed of light in a fixed direction then its centre of mass is effectively
a massless particle, and this is possible only if the string has a
vanishing mass-density.

It was shown in \cite{LM}, in the IIB D-string context, that the
total angular momentum of a 1/4-supersymmetric momentum-carrying string is
maximized, for fixed linear momentum, when the transverse profile is a
circle in a plane in $\bbe{8}$. Here we provide a much simpler proof of
this result. It turns out that there is a sequence of four bounds on the
angular momentum (given 1/4 supersymmetry), according to whether the
angular-momentum 2-form $L$ has rank 2, 4, 6 or 8. The `rank $2k$'
bound is saturated by curves in $\bbe{9}$ for which the projection onto
$\bbe{8}$ is a closed curve that has projections to circles of common
radii on $k$ orthogonal planes. These could be called
`generalized' superhelices; the $k=1$ case is the IIA version of the IIB
D-string `superhelix' described in \cite{CO}. As the $k=1$ bound is the
weakest one, it yields the maximum angular momentum for fixed linear
momentum; this bound is the same as the one obtained for a D-string in
\cite{LM} and for a supertube in \cite{MT,EMT,MNT}.

One of the motivations for the present work was to gain an understanding
of the implications of quantum mechanics for supertubes. Given the
freedom of an arbitrary cross-sectional shape, one may wonder, for
example, whether there is a unique quantum state associated to each
classical shape. The TST-dual supercurve provides a simpler context in
which to pose such questions because any IIA superstring configuration
corresponds to a state of the {\it perturbative} quantum IIA
superstring. We analyze the angular momentum of the 1/4-supersymmetric
`BPS' states of IIA superstring theory and derive a quantum version of
the classical angular momentum bound, which is recovered as a
semi-classical limit. It turns out that (for fixed linear momentum) there
is a unique one-string state of the perturbative IIA superstring theory for
which the quantum bound is saturated, as one might expect from the fact
that the classical bound is saturated by a unique wave profile.

%%%%%%%%%%%%%%%%%%%%%%%%%%%%%%%%%%%%%%%%%%%%%%%%%%%%%%%%%%%%%%%%%%%%%%%%%%%%
%%%%%%%%%%%%%%%%%%%%%%%%%%%%%%%%%%%%%%%%%%%%%%%%%%%%%%%%%%%%%%%%%%%%%%%%%%%%
%%%%%%%%%%%%%%%%%%%%%%  SECTION 2 %%%%%%%%%%%%%%%%%%%%%%%%%%%%%%%%%%%%%%%%%%
%%%%%%%%%%%%%%%%%%%%%%%%%%%%%%%%%%%%%%%%%%%%%%%%%%%%%%%%%%%%%%%%%%%%%%%%%%%%
%%%%%%%%%%%%%%%%%%%%%%%%%%%%%%%%%%%%%%%%%%%%%%%%%%%%%%%%%%%%%%%%%%%%%%%%%%%%
%\newpage
\sect{IIA Supercurves}
\label{iia}
%%%%%%%%%%%%%%%%%%%%%%%%%%%%%%%%%%%%%%%%%%%%%%%%%%%%%%%%%%%%%%%%%%%%%%%%%%%%

Let $\xi^a =(t,\sigma)$ be the string worldsheet coordinates and
$X^\mu=(X^0,\mbf{X})$ be Cartesian coordinates for the $D=10$ Minkowski
spacetime. We shall fix the time reparametrization invariance by the
choice
\be
X^0=t\, .
\ee
In this gauge, the induced worldsheet metric $g_{ab}$ is such that
\be
\sqrt{-\det g} = \sqrt{(1-\dot{\mbf{X}}^2)(\mbf{X}')^2 +
(\dot{\mbf{X}} \cdot \mbf{X}')^2}\, ,
\ee
where the overdot and prime indicate differentiation with respect to $t$
and $\sigma$, respectively. Let us now write $\mbf{X}=(Z,\mbf{Y})$
and set
\be\label{dots}
\dot Z=1\, ,\qquad \dot{\mbf{Y}} =\mbf{0}\, ,
\ee
in which case
\be\label{area}
\sqrt{-\det g} = |Z'|\, .
\ee
If we interpret the $Z$-direction as a T-dual direction then
configurations of this type are precisely what is obtained by a TST-duality
transformation of the  D2-brane supertube. The condition $\dot Z=1$,
which means that the string moves with the speed of light in the
$Z$-direction, arises from the condition that the BI
electric field on the supertube take the value that would be
critical in the absence of the magnetic field. Also, the condition
that the BI magnetic field not change sign translates to the same
condition on $Z'$; we shall see below that this is needed for
preservation of supersymmetry.

The condition for a bosonic IIA superstring configuration to preserve
some fraction of supersymmetry of the IIA Minkowski vacuum is that the
equation 
\be
\Gamma\epsilon =\epsilon
\label{susy}
\ee
admits non-zero constant spinor solutions $\epsilon$, where $\Gamma$
is the `kappa-symmetry' matrix
\be
\Gamma = {1\over \sqrt{-\det g}} \dot X^\mu X'^\nu
\Gamma_{\mu\nu}\Gamma_\natural\, .
\ee
As usual, $\Gamma_\natural$ denotes the product of all ten Dirac
matrices, anticommuting with each $\Gamma_\mu$ and squaring to the
identity. Note that $\Gamma^2=1$.
For configurations satisfying (\ref{dots}) we have
\be
\Gamma = {\rm sgn}(Z') \Gamma_{TZ}\Gamma_\natural  -
{1\over |Z'|}\, \mbf{Y}' \cdot \mbf{\Gamma}_Y \Gamma_T\Gamma_\natural
\left(1 - \Gamma_{TZ}\right).
\ee

Let us first consider the case in which $\mbf{Y}'/Z'$ is a constant
8-vector. As long as $Z'$ is everywhere non-zero, this corresponds to an
infinite straight string at an angle $\arctan (|\mbf{Y}'|/Z')$ to the
$Z$-axis. Although this string moves at the speed of light in the
$Z$-direction, only the orthogonal component is physical and this has the
subluminal magnitude
\be
v= {|\mbf{Y}'| \over \sqrt{(Z')^2 + |\mbf{Y}'|^2}}\, .
\ee
This case therefore corresponds to an orthogonally boosted infinite
straight string, which we would expect to preserve 1/2 supersymmetry. It
does so because in this case $\Gamma$ is a {\it constant} matrix
and the equation \eqn{susy} admits 16 linearly independent solutions
for $\epsilon$.

If $\mbf{Y}'/Z'$ is {\it not} a constant 8-vector, but $Z'$ is nowhere
zero, then the equation \eqn{susy} is solved by spinors satisfying
\be\label{spincon}
\Gamma_{TZ}\Gamma_\natural \, \epsilon = {\rm sgn}(Z') \, \epsilon
\sac \Gamma_{TZ} \, \epsilon = \epsilon \,.
\ee
These two conditions are those of a 1/4-supersymmetric superposition of a
string and linear momentum along the $Z$-axis, although the string now
describes an arbitrary curve in the `transverse' $\bbe{8}$ space.
These are the IIA supercurves that are TST-dual to the general
supertube, although one might wish to restrict the term `supertube'
to apply only to those cases for which the cross-section is
a closed non-intersecting curve in $\bbe{8}$.
For such cases the supercurve will (i) be periodic and (ii) have
non-zero angular momentum; it can be interpreted as a 1/2-supersymmetric
string parallel to the $Z$-axis that has expanded to a
1/4-supersymmetric supercurve as a result of its angular momentum.
There is an entirely analogous result for D-strings that generalizes
the super D-helix of \cite{CO}, which is the T-dual of the
circularly-symmetric supertube.

%%%%%%%%%%%%%%%%%%%%%%%%%%%%%%%%%%%%%%%%%%%%%%%%%%%%%%%%%%%%%%%%%%%%%%%%%%%%
%%%%%%%%%%%%%%%%%%%%%%%%%%%%%%%%%%%%%%%%%%%%%%%%%%%%%%%%%%%%%%%%%%%%%%%%%%%%
%%%%%%%%%%%%%%%%%%%%%% SECTION 3 %%%%%%%%%%%%%%%%%%%%%%%%%%%%%%%%%%%%%%%%%%%
%%%%%%%%%%%%%%%%%%%%%%%%%%%%%%%%%%%%%%%%%%%%%%%%%%%%%%%%%%%%%%%%%%%%%%%%%%%%
%%%%%%%%%%%%%%%%%%%%%%%%%%%%%%%%%%%%%%%%%%%%%%%%%%%%%%%%%%%%%%%%%%%%%%%%%%%%
%\newpage
\sect{Hamiltonian Analysis}
\label{ham}
%%%%%%%%%%%%%%%%%%%%%%%%%%%%%%%%%%%%%%%%%%%%%%%%%%%%%%%%%%%%%%%%%%%%%%%%%%%%

So far we have found a class of IIA string configurations that preserve
1/4 supersymmetry, but we should still check that they solve the
Nambu-Goto equations. We will do this by considering how these solutions
look in phase space. This will also permit a simple demonstration that
a supercurve is equivalent to a wave-carrying string.  
The (bosonic) phase space Lagrangian density is
\be
{\cal L} = P \cdot\dot X - {1\over2} e\left[P^2 +
(X')^2\right] - s \, P\cdot X'\, ,
\ee
where $e$ and $s$ are Lagrange multipliers that impose the
Hamiltonian and reparametrization constraints respectively.
In the $X^0=t$ gauge this becomes
\be
{\cal L} = P_Z D_tZ + \mbf{P} \cdot D_t\mbf{Y} - {\cal H} \, ,
\ee
where $D_t$ is the reparametrization-covariant time derivative
\be
D_t = \partial_t -s \partial_\sigma \, ,
\ee
and ${\cal H}$ is the (bosonic) Hamiltonian density. Using the reparametrization
constraint,
\be\label{repacon}
P_Z Z' + \mbf{P}\cdot \mbf{Y}' =0\, ,
\ee
we can write the Hamiltonian density as
\be\label{hamden}
{\cal H} = \sqrt{(P_Z \mp Z')^2 + |\mbf{P} \mp \mbf{Y}'|^2 }\, .
\ee
For this Hamiltonian density, the $P_Z$ and $\mbf{P}$ equations of
motion are
\bea
\label{covdiveqs}
D_t Z &=& {\cal H}^{-1}\left(P_Z \mp Z'\right) \,, \nn
D_t \mbf{Y} &=& {\cal H}^{-1} \left(\mbf{P} \mp \mbf{Y}'\right) \,.
\eea

It is obvious from (\ref{hamden}) that, in the $Z'=1$ gauge,
the energy is minimized for fixed $P_Z$ when
\be
\label{BPS}
\mbf{P} = \pm \mbf{Y}'\, .
\ee
It may be verified that such configurations solve both the equations
(\ref{covdiveqs}) and the constraint (\ref{repacon}) provided that
\be
|D_t Z| = 1\, , \qquad D_t Y = 0\, .
\ee
For $s=0$ this implies $|\dot Z|=1$ and $\dot{\mbf{Y}}={\bf 0}$, so we
conclude that the supercurve configurations described earlier indeed solve
the Nambu-Goto equations, and that they are equivalent to phase space
configurations satisfying (\ref{BPS}). Given (\ref{BPS}), they are also
solved by $\dot Z=0$ but for a different value of $s$ and
with $\dot{\mbf{Y}}$ now non-zero, such that $\dot{\mbf{Y}}=\pm \mbf{Y}'$ in
the $Z'=1$ gauge. This is the more conventional description of a supercurve
as a string carrying a wave of arbitrary profile that moves at the speed of
light in one direction along the string. The two descriptions are
equivalent because in both cases the {\it covariant} time derivatives of
$Z$ and $\mbf{Y}$ are the same, and for either description a case-by-case
analysis of the signs leads to the conclusion that
\be
{\cal H} = |Z'| + |P_Z|\, .
\ee
The first term is the energy due to the string mass density, and the
second term is the energy due to the momentum-carrying wave on
it\footnote{Note that the reparametrization constraint implies that
$P_Z=0$ when $\mbf{Y}'= {\bf 0}$, as expected since a string with
$\mbf{Y}'={\bf 0}$ is parallel to the $Z$-axis.}.

The Hamiltonian density is the time-time component of the string
stress-energy tensor density
\be
\tau^{ab} = -\sqrt{-\det g} \, g^{ab} \,,
\ee
where $g^{ab}$ is the inverse of the induced worldsheet metric.
Observe that this is divergence-free in the gauge $Z'=1$ as a
consequence of the $X^0$ and $Z$ equations of motion.
For the supercurve one finds in this gauge that
\be
\tau^{ab} = \pmatrix{{\cal H} & - 1 \cr -1 &0} \,.
\ee
The off-diagonal components correspond to the presence
of the linear momentum density carried by the wave. The vanishing of the
string tension $-\tau^{\sigma\sigma}$ provides one explanation of why
an arbitrary fixed wave profile in $\bbe{8}$ is stable.

%%%%%%%%%%%%%%%%%%%%%%%%%%%%%%%%%%%%%%%%%%%%%%%%%%%%%%%%%%%%%%%%%%%%%%%%%%%%
%%%%%%%%%%%%%%%%%%%%%%%%%%%%%%%%%%%%%%%%%%%%%%%%%%%%%%%%%%%%%%%%%%%%%%%%%%%%
%%%%%%%%%%%%%%%%%%%%%% SECTION 4 %%%%%%%%%%%%%%%%%%%%%%%%%%%%%%%%%%%%%%%%%%%
%%%%%%%%%%%%%%%%%%%%%%%%%%%%%%%%%%%%%%%%%%%%%%%%%%%%%%%%%%%%%%%%%%%%%%%%%%%%
%%%%%%%%%%%%%%%%%%%%%%%%%%%%%%%%%%%%%%%%%%%%%%%%%%%%%%%%%%%%%%%%%%%%%%%%%%%%
%\newpage
\sect{Angular Momentum Bound}
\label{sec-angmom}
%%%%%%%%%%%%%%%%%%%%%%%%%%%%%%%%%%%%%%%%%%%%%%%%%%%%%%%%%%%%%%%%%%%%%%%%%%%%

Let us suppose that $\mbf{Y}(\sigma)$ is non-zero and periodic with
period $2\pi$. The angular-momentum 2-form per period has components
\be\label{angmombos}
L_{ij} = {1\over 2\pi} \oint d\sigma \left(Y_iP_j - Y_jP_i\right).
\ee
Given (\ref{BPS}), this becomes
\be
L_{ij} = \pm{1\over 2\pi} \oint d\sigma (Y_iY'_j - Y_jY'_i)\, ,
\label{eqn-def-L}
\ee
which is proportional to the area of the closed curve formed by the
projection of $\mbf{Y}(\sigma)$ over one period onto the $ij$-plane. Recall
that (\ref{BPS}) was derived by minimizing the energy in the $Z'=1$ gauge
for fixed $P_Z$. We learn from the reparametrization constraint that the
momentum per period in the $Z$-direction is
\be
|\Delta P_Z| = {1\over 2\pi}\oint d\sigma |\mbf{Y}'|^2
\label{constraint}
\ee
in the $Z'=1$ gauge. The integrand is the square of the length-density of
the curve $\mbf{Y}(\sigma)$, so we might expect, by some variant of
the isoperimetric inequality, that the total squared angular momentum per
period
\be
J^2 = {1\over2} L_{ij}L^{ij}
\ee
will be bounded by some multiple of $\Delta P_Z$. We shall now establish
this bound.

Specifically, we seek to maximize $J^2$ subject to the constraint
\eqn{constraint}. This is equivalent to maximizing, without
constraint, the functional
\be
\calf[\mbf{Y}; \lambda] = J^2[\mbf{Y}] + 2 \lambda \left[\oint
d\sigma |\mbf{Y}'|^2  \ - 2\pi |\Delta P_Z| \right] 
\ee
with respect to the functions $\mbf{Y}$ and the Lagrange multiplier
variable $\lambda$. Variation with respect to the functions $Y_i$
yields second-order differential equations which are trivially
once-integrated. Setting the integration constants to zero by a
translation in $\bbe{8}$ (under which $J$ is invariant) we arrive
at the equation
\be
\label{ELeq}
Y_i '  = \frac{1}{\lambda J} \, L_{ij} Y^j \,.
\ee
Contracting with $Y_i'$ and integrating over a period we deduce that
\be
\lambda = \frac{J}{|\Delta P_Z|} \,.
\label{lamb}
\ee

By a rotation in $\bbe{8}$ (under which $J$ is invariant)
we can skew-diagonalize $L$; let $\ell_p$, $p=1,\ldots ,4$,
be the skew-eigenvalues. $L$ is then diagonal (with eigenvalues
$\pm i \ell_p$) in the complex basis
\be
W_p = Y_{2p -1} + i Y_{2p} \,,
\ee
and the solution of (\ref{ELeq}) is then
\be
\label{solution}
W_p = R_p \, e^{iN_p\sigma} \,.
\ee
Here $R_p$ are arbitrary constants, and the integers
\be
\label{la}
N_p = \frac{|\Delta P_Z|}{J^2} \, \ell_p
\ee
count the number of times per period that the string winds around the
origin of the complex $W_p$-plane. Note that the overall phase of each
of the four complex functions $W_p$ in \eqn{solution} may be brought
to any desired value by a transformation in the residual $SO(2)^4$
subgroup of $SO(8)$ that preserves the skew-diagonal form of $L$.

Given \eqn{solution}, $L$ may be computed and one can thus verify that
it {\it is} skew-diagonal. This calculation also reveals that
\be
\ell_p = N_p \, R_p^2 \,.
\label{comparison}
\ee
If $\ell_p \neq 0$ then this together with \eqn{la} shows that
\be
R_p^2 = \frac{J^2}{2\pi |\Delta P_Z|} \,.
\label{Ra}
\ee
If $\ell_p=0$ then (\ref{la}) implies that $N_p=0$ too, so that
$R_p$ is undetermined and $W_p$ is constant; by a translation in the
complex $W_p$-plane we may still choose $R_p$ to be given by the
formula \eqn{Ra}. Thus, the solution (\ref{solution}) becomes
\be\label{genhel}
W_p = \sqrt{\frac{J^2}{2\pi |\Delta P_Z|}} \, e^{iN_p\sigma} \,.
\ee
Defining
\be
N^2 \equiv \sum_{p=1}^4 N_p^2 
\ee
and making use of \eqn{la}, we see that the angular momentum of this
solution saturates the bound
\be
J \leqslant N^{-1} \, |\Delta P_Z|
\ee
for fixed linear momentum per period $|\Delta P_Z|$ and winding
numbers $N_p$.

The geometry of the periodic supercurve with maximal angular momentum
per period therefore depends in an essential way on the rank of the
angular-momentum 2-form; through \eqn{la} this is determined by the
number of non-zero winding numbers $N_p$, which, of course, may be
specified
independently of the linear momentum. If only $|\Delta P_Z|$ is
specified, then the absolute bound on the angular momentum becomes
simply 
\be
\label{jbound}
J \leqslant |\Delta P_Z| \,,
\ee
which is the bound discussed in \cite{LM} in the ST-dual context of the
D-string and which is saturated when (say) $N_1=1$ and
$N_2=N_3=N_4=0$. In this case $L$ has rank 2 and the supercurve is a
helix (of fixed pitch in the $Z'=1$ gauge) in an $\bbe{3}$ subspace of
$\bbe{9}$; in fact, it is the IIA dual of the superhelix of \cite{CO}.

%%%%%%%%%%%%%%%%%%%%%%%%%%%%%%%%%%%%%%%%%%%%%%%%%%%%%%%%%%%%%%%%%%%%%%%%%%%%
%%%%%%%%%%%%%%%%%%%%%%%%%%%%%%%%%%%%%%%%%%%%%%%%%%%%%%%%%%%%%%%%%%%%%%%%%%%%
%%%%%%%%%%%%%%%%%%%%%% SECTION 5 %%%%%%%%%%%%%%%%%%%%%%%%%%%%%%%%%%%%%%%%%%%
%%%%%%%%%%%%%%%%%%%%%%%%%%%%%%%%%%%%%%%%%%%%%%%%%%%%%%%%%%%%%%%%%%%%%%%%%%%%
%%%%%%%%%%%%%%%%%%%%%%%%%%%%%%%%%%%%%%%%%%%%%%%%%%%%%%%%%%%%%%%%%%%%%%%%%%%%
%\newpage
\sect{Light-front Gauge}
\label{light-front}
%%%%%%%%%%%%%%%%%%%%%%%%%%%%%%%%%%%%%%%%%%%%%%%%%%%%%%%%%%%%%%%%%%%%%%%%%%%%

With a view to simplifying the quantum treatment of supercurves within the
context of type II superstring theory, we now reobtain some of the
above results in the light-front gauge. We define
\be
X^\pm = \fc{X^0 \pm Z}{\sqrt{2}} \sac
P_\pm = \fc{P_0 \pm P_Z}{\sqrt{2}}
\ee
and set 
\be
X^+ =t \sac P_- = -p^+
\ee
for positive constant $p^+$. The Hamiltonian constraint may now be solved
for the light-front Hamiltonian density ${\cal H}=-P_+$, while
the reparametrization constraint implies that
\be\label{repa}
(X^-)' = {\mbf{P}\cdot \mbf{Y}' \over p^+}\, .
\ee
The bosonic Lagrangian density becomes
\be
\label{bos}
{\cal L}_{bos} = \mbf{P} \cdot \dot{\mbf{Y}} - {1\over
2p^+}\left(|\mbf{P}|^2 + |\mbf{Y}'|^2\right) - \partial_t\left(p^+
X^-\right)\, ,
\ee
from which one sees that
\be
\label{lfp}
\mbf{P} = p^+ \dot{\mbf{Y}}\, .
\ee

Using (\ref{repa}), we may rewrite the light-front gauge
bosonic Hamiltonian density as
\be
{\cal H}_{bos} = {1\over2p^+} |\mbf{P} \mp \mbf{Y}'|^2 \pm \partial_\sigma
X^-\, .
\ee
For periodic $\mbf{Y}$ the integral of ${\cal H}_{bos}$ over one period is
minimized, for fixed increase $\Delta X^-$ of $X^-$ over one period, when
\be
\label{lfbps}
\mbf{P} = \pm  \mbf{Y}'\, .
\ee
Given this relation one can show that
\be\label{Xmin}
(X^-)' = \pm {|\mbf{Y}'|^2 \over p^+}\, ,\qquad
\dot X^- = {|\mbf{Y}'|^2\over (p^+)^2}\, ,
\ee
and hence that 
\be
\Delta X^-= \pm {1\over p^+} \oint d\sigma\, |\mbf{Y}'|^2 \,,
\ee
where the integral is over one period. As $(X^-)' = -\sqrt{2} Z'$,
a string with zero $\mbf{Y}'$ also has zero $Z'$ and is therefore
pointlike. Conversely, if $\mbf{Y}'$ is non-zero then so is $\Delta Z$,
so we have a periodic wave along an infinite string in the $Z$-direction.

To determine the fraction of supersymmetry preserved by solutions satisfying
(\ref{lfbps}), we return to the supersymmetry preservation condition
(\ref{susy}). Imposing the conditions
\be
\Gamma_+\epsilon=0\, ,\qquad \Gamma_\natural \epsilon= \pm \epsilon\, ,
\ee
which are equivalent to (\ref{spincon}), we find that (\ref{susy})
becomes an identity when (\ref{lfbps}) is satisfied, by virtue of the
relations (\ref{repa}), (\ref{Xmin}) and (\ref{lfp}). Thus, as expected,
1/4 supersymmetry is preserved.

The full IIA superstring action in the light-front gauge of course
includes the Green-Schwarz worldsheet fermions $S^\pm$ in the
${\bf 8}_s$ and ${\bf 8}_c$ spinor representations of $SO(8)$. The
phase-space Lagrangian density generalizing (\ref{bos}) is (omitting
total derivatives)
\be
{\cal L} = \mbf{P} \cdot \dot{\mbf{Y}} -i(S^+)^T \dot S^+ -i
(S^-)^T\dot S^- - {\cal H} \,,
\ee
where the Hamiltonian density is
\be
{\cal H} = {1\over 2p^+} \left[ |\mbf{P}|^2 + |\mbf{Y}'|^2 -i (S^+)^T
(S^+)' +i (S^-)^T (S^-)'\right]\, .
\ee
The angular-momentum 2-form generalizing the bosonic expression of
(\ref{angmombos}) is
\be
L_{ij} = {1\over 2\pi} \oint d\sigma \left[ Y_iP_j - Y_jP_i
- {1\over 4} (S^+)^T \gamma_{ij} \, S^+ - {1\over 4} (S^-)^T
\gamma_{ij} \, S^- \right] \,,
\ee
where $\gamma_{ij}$ is the antisymmetrized product of a pair of $SO(8)$
Dirac matrices. 

The phase space action is invariant under the supersymmetry
transformations generated by
\be
Q_\pm = {1\over 2\pi \sqrt{2p^+}}\oint d\sigma \,
(\mbf{P} \pm \mbf{Y}') \cdot \mbf{\gamma} \, S^\mp \, .
\ee
We note here for future use that 
\be
K = -{1\over2\pi}\oint d\sigma \, \left[
\mbf{P} \cdot \mbf{Y}' -
{i\over2} (S^+)^T (S^+)' - {i\over2} (S^-)^T (S^-)' \right]
\ee
is invariant under these transformations. When $\mbf{P}+ \mbf{Y}' =0$,
the functional $K$ becomes the supersymmetric  extension of the
right-hand side of \eqn{constraint}, and hence equal to the quantity
$|\Delta P_Z|$ held fixed in deriving the classical angular momentum
bound. The fermions make no difference classically but we shall need
to consider them in the following discussion of the
quantum-mechanical angular momentum bound.

%%%%%%%%%%%%%%%%%%%%%%%%%%%%%%%%%%%%%%%%%%%%%%%%%%%%%%%%%%%%%%%%%%%%%%%%%%%%
%%%%%%%%%%%%%%%%%%%%%%%%%%%%%%%%%%%%%%%%%%%%%%%%%%%%%%%%%%%%%%%%%%%%%%%%%%%%
%%%%%%%%%%%%%%%%%%%%%% SECTION 6 %%%%%%%%%%%%%%%%%%%%%%%%%%%%%%%%%%%%%%%%%%%
%%%%%%%%%%%%%%%%%%%%%%%%%%%%%%%%%%%%%%%%%%%%%%%%%%%%%%%%%%%%%%%%%%%%%%%%%%%%
%%%%%%%%%%%%%%%%%%%%%%%%%%%%%%%%%%%%%%%%%%%%%%%%%%%%%%%%%%%%%%%%%%%%%%%%%%%%
%\newpage
\sect{Quantum Mechanical Bound}
\label{QM}
%%%%%%%%%%%%%%%%%%%%%%%%%%%%%%%%%%%%%%%%%%%%%%%%%%%%%%%%%%%%%%%%%%%%%%%%%%%%

The worldsheet fields can be expressed as Fourier series, the
coefficients of which become annihilation and creation operators in the
quantum theory. In particular, for zero total transverse momentum,
we have the oscillator expansions
\bea
\mbf{P} \mp \mbf{Y}' &=& 2\sum_{n=1}^\infty \sqrt{n} \left[
\mbf{a}_n^{\pm} e^{\pm in\sigma} + (\mbf{a}_n^{\pm})^\dagger
e^{\mp in\sigma}\right] \,, \nn
S^\pm &=& s^\pm + \sum_{n=1}^\infty \left[S_n^{\pm} e^{\pm in\sigma} +
(S_n^{\pm})^\dagger e^{\mp in\sigma} \right] \,,
\eea
where $\mbf{a}_n^{\pm}$ and $S_n^{\pm}$ are, respectively, the boson and
fermion oscillator annihilation operators. It follows that
\be
Q_\pm = \sqrt{\frac{2}{p^+}} \, \sum_{n=1}^\infty \sqrt{n} \,
\left[ \mbf{a}_n^{\pm} \cdot \mbf{\gamma}  \, (S_n^{\pm})^\dagger +
(\mbf{a}_n^{\pm})^\dagger \cdot \mbf{\gamma} \, S_n^{\pm} \right] \,,
\ee
and also that
\be
K = \sum_{n=1}^\infty n \, \left( \caln_n^B + \caln_n^F \right) \,,
\ee
where $\caln_n^B$ and $\caln_n^F$ are, respectively, the boson and
fermion number operators for the oscillators of the $n$th Fourier mode.

One-string states are built by the action of the creation operators on the
oscillator vacuum 
\be\label{oscvac}
|0\rangle = |0\rangle_+\otimes |0\rangle_- \,,
\ee 
which is annihilated by both $Q_+$ and $Q_-$. However, because of
the fermion zero modes $s^\pm$, the one-string vacuum is actually the
tensor product of  an ${\bf 8}_v\oplus {\bf 8}_c$ $SO(8)$ multiplet of
$|0\rangle_+$ vacua with an ${\bf 8}_v\oplus {\bf 8}_s$ multiplet of
$|0\rangle_-$ vacua.  The bosonic vacua are thus in the
\be\label{decomp}
{\bf 1} \oplus {\bf 8} \oplus {\bf 28} \oplus {\bf 35} \oplus {\bf 56}
\ee
tensor representation of $SO(8)$. Initially, at least, we will
consider only the
$SO(8)$ singlet because we can then ignore the effects of the fermion
zero modes on the angular momentum. Thus we may take the one-string vacuum
to be the oscillator vacuum (\ref{oscvac}). The one-string states of the
form
\be
\label{states}
|\psi\rangle = |\psi\rangle_+ \otimes |0\rangle_- 
\ee
with $|\psi\rangle_+ \ne |0\rangle_+$ are annihilated by $Q_-$ but not by
$Q_+$, and hence preserve 1/4 supersymmetry. These, and the analogous
states annihilated by $Q_+$ (but not by $Q_-$) are, as is well-known, the
1/4-supersymmetric `BPS' states of the quantum IIA superstring theory (in
light-front gauge).

Henceforth we shall need to consider only the `$+$' oscillators so we
omit the `$+$' superscript. A basis for the $|\psi\rangle_+$ factor
in \eqn{states} is 
\be
\label{basis2}
\bigotimes_{n=1}^\infty |B\rangle_n \otimes |F\rangle_n\,,
\ee
where
\be
|B\rangle_n = 
\prod_{i=1}^8 \, [(a_n^i)^\dagger]^{B_n^i} \, |0\rangle_n^B \sac
|F\rangle_n = 
\prod_{\alpha=1}^8 \, [(S_n^\alpha)^\dagger]^{F_n^\alpha}\,
|0\rangle_n^F \,.
\label{subbasis}
\ee
Here $B_n^i$ and $F_n^\alpha$ are non-negative integers, and
$|0\rangle_n^B$ and $|0\rangle_n^F$ are the ground states of the $n$th
bosonic and fermionic oscillators, respectively. Note that each of
the states \eqn{basis2} is an eigenstate of the 
bosonic and fermionic number operators for the $n$th oscillator with
eigenvalues
\be
B_n = \sum_{i=1}^8 N_n^i \sac F_n = \sum_{\a =1}^8 F_n^\a \,.
\ee
Bosonic BPS states are the subset of these states for which
$\sum_{n=1}^\infty F_n$ is even (since we chose to construct them on a
bosonic one-string vacuum).

Given the oscillator expansion of the worldsheet fields, the angular
momentum 2-form $L_{ij}$ is now the operator
\be
L_{ij} = \sum_{n=1}^\infty \left[ 2i \, (a_n^\dagger)_{[i}
(a_n)_{j]}
\nn -{1\over2} S_n^\dagger \gamma_{ij} S_n \right] + \cdots \,,
\ee
where we have omitted the zero mode contribution, and the ellipsis
indicates terms involving the `$-$' oscillators that are irrelevant to
matrix elements involving only the BPS states. Since $L_{ij}$ commute
with the oscillator number operators, states of the form
(\ref{subbasis}) for fixed $B_n$ and $F_n$ can be assembled into $SO(8)$
multiplets. Each of these  multiplets may be characterized by its Dynkin
labels  $\mbf{w}=(w^1, w^2, w^3, w^4)$, in terms of which the quadratic
Casimir of $SO(8)$ takes the form \cite{Slansky}
\be\label{jsquared}
J^2 = \sum_{a,b =1}^4 \, G_{ab} \, w^a \, (w^b +2) \,,
\ee
where 
\be
G_{ab} =  \pmatrix { 1   & 1   & 1/2  & 1/2  \cr
                     1   & 2   &  1   &  1   \cr
                    1/2  & 1   &  1   & 1/2  \cr
                    1/2  & 1   & 1/2  &  1   \cr} \,.
\ee
Note that the formula (\ref{jsquared}) is independent of $n$, whereas
the expression for $K$ is such that the higher the level of an
oscillator, the larger is its contribution to $K$.

We seek states that maximize $J^2$ for fixed $K$, or equivalently
states that minimize $K$ for fixed $J^2$. To this end let us write
$K$ in terms of the separate boson and fermion contributions as
$K=K_B + K_F$. Suppose for the moment that $K_F=0$; in this case it
is clear that the angular momentum will be maximized for fixed $K$
when $B_n=0$ for $n>1$, and hence $B_1=K$, because we may otherwise reduce
$K$ without changing $J^2$ by the substitution $n\rightarrow 1$.
This argument fails for the fermi oscillators because $F_n \leqslant 8$,
for all $n$, due to the Pauli exclusion principle. Thus, once we include
fermi oscillators it is no longer obvious that maximizing $J^2$
for fixed $K$ requires all oscillators with $n>1$ to be in their ground
states. Nevertheless, it seems to be true for small values of $K$ and in
these cases {\it all} fermion oscillators are in their ground states.
An obvious conjecture is that this will remain true for all $K$. We will
examine the consequences of this conjecture below but we should first
stress that it is not needed in the semi-classical limit of large $K$,
for the following reason. If $K$ is large then either $K_B$ or $K_F$, or
both, must be large. But if $K_F$ is large then most of the fermion
oscillators must be in levels with $n \gg 1$ because of the
exclusion principle, so a large number of fermions would make a large
contribution to $K$ for correspondingly small contribution to $J^2$ (as
compared to boson oscillators). Thus, for large $K$, the angular
momentum will be maximized in some state for which $K_B \gg K_F$,
and this means that the fermion oscillators contribute negligibly to the
angular momentum in the semi-classical limit; we may then assume, for
simplicity, that all fermion oscillators are in their ground states.
Note that the effect of choosing a non-singlet $SO(8)$ irrep in the
decomposition (\ref{decomp}) also has a negligible effect in the limit of
large $J$, thus justifying our simplifying assumption that the one-string
vacuum is an $SO(8)$ singlet.

Since the state with maximal $J$ for fixed $K$ has all bosonic
oscillators with $n>1$ in their ground states, the assumption that all
fermion oscillators are in their ground states leaves only BPS states of
the form
\be
\prod_i [(a_1^i)^\dagger]^K |0\rangle_+ \otimes |0\rangle_- \,.
\ee
These states transform in the $[(\mbf{8_v})^K]_\mathrm{sym}$
representation of $SO(8)$.  Within this reducible representation, the
$SO(8)$ irrep corresponding to the traceless $K$th rank tensor has the
largest value of $J^2$; this irrep has Dynkin labels
\be
\mbf{w} = (K, 0, 0, 0) \,.
\label{rep}
\ee
Using (\ref{jsquared}) we deduce that $J^2= K^2 +6K$ for this
representation, and hence the bound
\be
J^2 \leqslant  K^2 + 6 K \,,
\ee
with equality for the representation \eqn{rep}. Depending on the validity
of the conjecture above, and the effect of choosing a non-singlet
representation in (\ref{decomp}), this bound may require adjustment for
small $K$, but it is certainly valid in the semi-classical limit of large
$K$, for which it reduces to $J\leqslant K$. We have derived this bound in the
light-front gauge but the result can of course be applied in any gauge;
in particular in the `physical' gauge $\dot X^0=Z'=1$, in which $K$
equals $|\Delta P_Z|$. We thus recover the classical bound \eqn{jbound}.

%%%%%%%%%%%%%%%%%%%%%%%%%%%%%%%%%%%%%%%%%%%%%%%%%%%%%%%%%%%%%%%%%%%%%%%%%%%%
%%%%%%%%%%%%%%%%%%%%%%%%%%%%%%%%%%%%%%%%%%%%%%%%%%%%%%%%%%%%%%%%%%%%%%%%%%%%
%%%%%%%%%%%%%%%%%%%%%% SECTION 7 %%%%%%%%%%%%%%%%%%%%%%%%%%%%%%%%%%%%%%%%%%%
%%%%%%%%%%%%%%%%%%%%%%%%%%%%%%%%%%%%%%%%%%%%%%%%%%%%%%%%%%%%%%%%%%%%%%%%%%%%
%%%%%%%%%%%%%%%%%%%%%%%%%%%%%%%%%%%%%%%%%%%%%%%%%%%%%%%%%%%%%%%%%%%%%%%%%%%%
%\newpage
\sect{Conclusions}
\label{conc}
%%%%%%%%%%%%%%%%%%%%%%%%%%%%%%%%%%%%%%%%%%%%%%%%%%%%%%%%%%%%%%%%%%%%%%%%%%%%

This paper was motivated by a desire to better understand the properties
of D2-brane supertubes of general cross section by considering the
TST-dual IIA superstring configuration. This is a 1/4-supersymmetric IIA
string carrying both momentum and angular momentum that we have called a
`supercurve'; it includes as a special case the `superhelix' described
for D-strings in \cite{CO}. In fact, IIA supercurves just provide an
alternative description of a IIA string carrying a left or right moving
wave of arbitrary profile. As it has long been known that such
configurations preserve 1/4 supersymmetry, this may help demystify the
result of \cite{MNT} that 1/4 supersymmetry allows a supertube to have
an arbitrary cross-section in $\bbe{8}$.

The main focus of this paper has been on the angular momentum carried by
supercurves. The angular momentum of a supertube is subject to an upper
bound that is saturated by supertubes of planar and circular
cross-section \cite{MT,EMT,MNT}. The latter have superhelices as their TST
duals, so one would expect superhelices to saturate an upper bound on
the angular momentum (for fixed linear momentum). A
classical bound of this type was derived in \cite{LM} in a slightly
different context. We have given a much simpler proof of it, and a
refinement to the case in which one fixes the rank of the angular
momentum 2-form in addition to the linear momentum.

We have shown how the classical bound on the angular momentum can be
recovered from the semi-classical limit of a quantum bound derived within
the context of perturbative IIA superstring theory. There is a unique
$SO(8)$ representation for which the angular momentum is maximal.
We would certainly expect that many of the properties of quantum
supercurves continue to hold for quantum supertubes, and that was
another motivation for their study. Our results lead us to expect that
a quantum supertube of maximal angular momentum will again be associated
to a unique $SO(8)$ representation.

%%%%%%%%%%%%%%%%%%%%%%%%%%%%%%%%%%%%%%%%%%%%%%%%%%%%%%%%%%%%%%%%%%%%%%%%%%%
%%%%%%%%%%%%%%%%%%%%%%%%%%%%%%%%%%%%%%%%%%%%%%%%%%%%%%%%%%%%%%%%%%%%%%%%%%%
%%%%%%%%%%%%%%%%%%%%%% ACKNOWLEDGMENTS %%%%%%%%%%%%%%%%%%%%%%%%%%%%%%%%%%%%
%%%%%%%%%%%%%%%%%%%%%%%%%%%%%%%%%%%%%%%%%%%%%%%%%%%%%%%%%%%%%%%%%%%%%%%%%%%
%%%%%%%%%%%%%%%%%%%%%%%%%%%%%%%%%%%%%%%%%%%%%%%%%%%%%%%%%%%%%%%%%%%%%%%%%%%
%\newpage
\section*{Acknowledgments}
%%%%%%%%%%%%%%%%%%%%%%%%%%%%%%%%%%%%%%%%%%%%%%%%%%%%%%%%%%%%%%%%%%%%%%%%%%%

We are grateful to Jerome Gauntlett, Jaume Gomis,
Joaquim Gomis and Daniel Waldram for
helpful discussions. D.M. is supported by a PPARC fellowship.  S.N. is
supported by the British Federation of Women Graduates and the
Australian Federation of University Women (Queensland).

%%%%%%%%%%%%%%%%%%%%%%%%%%%%%%%%%%%%%%%%%%%%%%%%%%%%%%%%%%%%%%%%%%%%%%%%%%%
%%%%%%%%%%%%%%%%%%%%%%%%%%%%%%%%%%%%%%%%%%%%%%%%%%%%%%%%%%%%%%%%%%%%%%%%%%%
%%%%%%%%%%%%%%%%%%%%%% BIBLIOGRAPHY %% %%%%%%%%%%%%%%%%%%%%%%%%%%%%%%%%%%%%
%%%%%%%%%%%%%%%%%%%%%%%%%%%%%%%%%%%%%%%%%%%%%%%%%%%%%%%%%%%%%%%%%%%%%%%%%%%
%%%%%%%%%%%%%%%%%%%%%%%%%%%%%%%%%%%%%%%%%%%%%%%%%%%%%%%%%%%%%%%%%%%%%%%%%%%

%%%%%%%%%%%%%%%%%%%%%%%%%%%%%%%%%%%%%%%%%%%%%%%%%%%%%%%%%%%%%%%%%%%%%%%%%

\end{document}